# Counting and differentiating aquatic biotic nanoparticles by full-field interferometry: from laboratory tests to *Tara* Oceans sample analysis

**Martine Boccara**[1,2,#,*], **Yasmina Fedala**[1,3,#], **Catherine Venien Bryan**[4], **Marc Bailly-Bechet**[2,5], **Chris Bowler**[1] **and Albert Claude Boccara**[3,*]

[1] *Ecole Normale Supérieure, PSL Research University, Institut de Biologie de l'Ecole Normale Supérieure (IBENS), CNRS UMR 8197, INSERM U1024, 46 rue d'Ulm, F-75005 Paris, France.*
[2] *Atelier de Bioinformatique, UMR 7205 ISYEB CNRS, MNHN, UPMC, EPHE, 45 rue Buffon, 75005 Paris, France.*
[3] *Institut Langevin, ESPCI ParisTech, PSL Research University, CNRS UMR 7587, 1 rue Jussieu, 75005 Paris, France.*
[4] *Institut de Minéralogie et de Physique des Milieux Condensés  UMR 7590, UPMC, Paris, France.*
[5] *Laboratoire Biométrie et Biologie Evolutive, Université Claude Bernard Lyon 1, CNRS, UMR5558, Bâtiment Gregor Mendel, 43 Boulevard du 11 Novembre 1918, 69622 Villeurbanne, France.*

# These authors contributed equally to this paper

\* mboccara@biologie.ens.fr, claude.boccara@espci.fr

**Abstract:** There is a huge abundance of viruses and membrane vesicles in seawater. We describe a new full-field, incoherently illuminated, shot-noise limited, common-path interferometric detection method that we couple with the analysis of Brownian motion to detect, quantify, and differentiate biotic nanoparticles. We validated the method with calibrated nanoparticles and homogeneous DNA or RNA.viruses. The smallest virus size that we characterized with a suitable signal-to-noise ratio was around 30 nm in diameter. Analysis of Brownian motions revealed anisotropic trajectories for myoviruses.We further applied the method for vesicles detection and for analysis of coastal and oligotrophic samples from *Tara* Oceans circumnavigation.

**1. Introduction**
Marine environments are rich in nanoparticles such as viruses of various origins and composition [1-6]; membrane vesicles are another type of nanoparticles, they contained proteins and lipids, as well as DNA and RNA and recent evidence has shown that they are produced in the open ocean [7,8]. It is thus of primary importance to enumerate and distinguish these different types of biotic nanoparticles in marine as well as in other ecosystems**.**
To date, most attention has focused on quantifying viral particles. Optical methods based on epifluorescence microscopy to detect dsDNA binding fluorophores such as DAPI or SYBR green have also been used extensively [9,10,11] but are not adapted for the detection of single stranded DNA viruses, nor RNA-containing viruses [12,13]. Furthermore, although epifluorescence can enumerate membrane vesicles containing genetic material they cannot detect empty vesicles [14].
Methods based on light scattering by suspensions of *identical* viral particles have been mostly applied to homogeneous virus suspensions [15]. A particle smaller than the resolution will appear under microscopic observation as a diffraction-limited spot that exhibits intensity proportional to the amount of scattered light. Because capsid-encased viruses behave as nanometer-sized dielectric particles with refractive indexes close to 1.5 [16], one can compute their scattering cross sections and the amount of scattered power. A resonant amplification of this weak scattered power level can be obtained when viruses are stuck to metallic [17,18] or dielectric structures [19,20]. Optical homodyne or heterodyne detections increases signal through the use of a local oscillator [21]; in this latter approach a *single* detector was used that counts single particles successively.

An additional methodology to track nanoparticles takes advantage of their individual Brownian motions in an imaging microscope [22,23]: nanoparticle tracking analysis (NTA) is one such method based on detection of Brownian motion of supposed spherical particles as a function of their size [24]. However, Brownian motion based detection cannot differentiate vesicles from viruses of the same size.

Because of the limitations of current technologies to comprehensively differentiate and quantify viruses and vesicles we aimed to develop a method able to detect and quantify different types of biotic nanoparticles in environmental samples. Here we present a method for successful viral identification and counting over a broad range of sizes and shapes using a new full-field sensitive imaging method that takes advantage of the interference phenomena between a strong optical reference signal and the weak power scattered by particles to amplify the scattered signal. With this device, we measure for each particle their scattering signal and follow their Brownian motion trajectories. We validate this new method with calibrated nanoparticles and well-characterized representative viruses (30-100 nm range), and were also able to detect and enumerate vesicles secreted by the marine diatom *Phaeodactylum tricornutum*. We further show that vesicles can be differentiated from viruses and we analyze samples from contrasting marine environments collected during the *Tara* oceans circumnavigation [25,26].

## 2. Instrument design and validation with 50 nm beads

In order to simplify full-field interferometric systems [27,28], we developed a very simple and stable common path system (**Figs. 1A and 1B**) that appears to be well adapted to the parallel detection of nanoparticles. For each diffraction limited spot of the incoming light field (radius $\rho$, wavelength $\lambda$ linked to the numerical aperture (NA) of the objective ($\rho=1.22\ \lambda/2NA$) that impinges the particle, the interference takes place near the focus, between the strong transmitted beam and the weak forward scattered one. Because the incoming and the scattered beams follow the same path there is no other phase shift than the Gouy phase shift (**Fig. 1C**) that happens close to the focus and makes the interferences destructive or constructive [29].

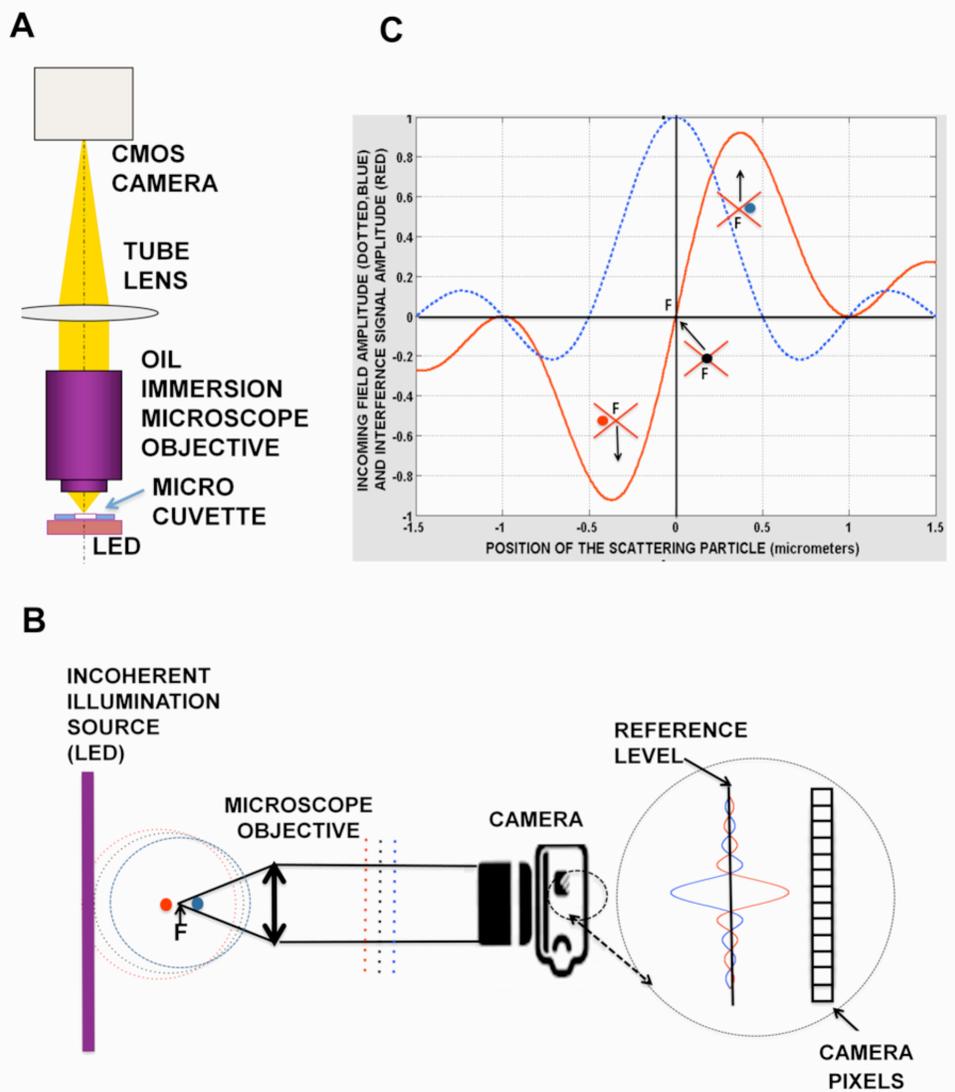

Fig.1. Principle of the interferometry method. A) Experimental set up: The sample to be analyzed fills the micro-cuvette (in white) bounded by coverslips and located at the focus of the oil immersion objective. A bare LED illuminates the cuvette. The common path interferometer uses the incoherent flux of a LED for a light source and a single microscope objective whose focal plane is imaged on the camera CMOS chip. B) Principle of the common path detection. The sample is illuminated by an incoherent source, so the spatial coherence expands over a diffraction spot whose size is linked to the numerical aperture of the microscope objective. The illuminating wave, restricted to this diffraction spot, is able to interfere with the light scattered by the particles that are located close (within the depth of field) to the objective focus F. Due to the Gouy phase shift the interference is constructive or destructive for a particle located slightly after (blue dot and dotted line) or before (red dot and dotted line) the focus. C) Calculated signal amplitude (in arbitrary units) as a function of the distance of the scattering particle to the focus F of the microscope objective (red crosses). The amplitude of the incoming field is the blue dotted line. At the focus the scattered particle field is in quadrature with the incoming wave so the interference signal is zero (black dot and dotted line).

We can first estimate the sensitivity of our set up. Using the full well capacity of our camera (>160 000 e$^-$) and a standard signal spatial filtering (see below), the noise equivalent scattered signal was found to be $10^{-7}$ of the incident power. We can then compute the minimum scattering cross section $\sigma_{min}$ that could be detected from this value: for a collection efficiency of 0.25 (solid angle compared to 4π, quantum efficiency of the camera and transmission of the optics) and a diffraction spot surface of about 0.044 µm$^2$, we estimated $\sigma_{min} \simeq 10^{-8}$ µm$^2$. The scattering cross sections $\sigma$ at 450nm wavelength (Mie scattering calculator, http://omlc.ogi.edu/calc/mie_calc.html) for particles of diameters of 100, 70, 50, 35 and 20 nm of refractive index 1.5 (n) in water are, respectively, 8 10$^{-5}$, 1.1 10$^{-5}$, 1.5 10$^{-6}$ 1.8 10$^{-7}$ and 0.7 10$^{-8}$ µm$^2$. These values confirm our experimental observation where the noise level corresponds to the light scattered by a particle diameter of around 20 nm.

For our measurements the sample (about 5 µL) is deposited in a 100 µm thick micro-cuvette in which convection induced drifts and damping of the light power by absorption are unlikely. We first calibrated our settings using 50 nm diameter silica beads. The field of view recorded by the camera is 80 x 80 µm$^2$. We intentionally oversampled the diffraction spot (its diameter corresponds to 10 pixels) in order to increase the signal-to-noise-ratio through a normalized convolution by the diffraction spot **(Fig. 2)**. Indeed the Bessel function describing the diffraction spot is invariant by convolution with itself; if this function is normalized the signal remains unchanged while the noise is reduced about 4-fold **(Fig. 2)**.

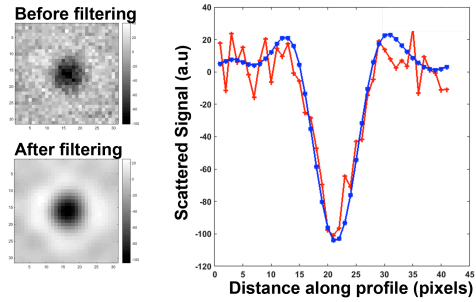

Fig. 2. Signal Processing. For a given spot the intensity of the scattered signal was measured after subtracting from the raw image the average of the stack of 200 images (before filtering, red curve). The normalized convolution then reduces the noise without affecting the signal (after filtering, blue curve). On the left images before and after filtering

We follow the trajectory of each particle as a function of time. When recording the time varying *amplitude* associated with the displacement of the scattered signal we detect the maximum and/or the minimum signal.. To follow the Brownian motion of each particle we take a movie of the field of view over a few seconds (typically around 2 seconds with the 150 Hz New Focus CMOS camera). We measured the signal from 50 nm calibrated beads (diluted to obtain a concentration of $10^{10}$ particles/mL) with a refractive index close to 1.47 at the wavelength of the source (450nm). We observed a single peaked histogram, which we centered at 50 nm with a scattering signal level of 30 ± 10 a.u (SD) **(Figs. S1A, S1B)**. The distribution of sizes values observed by our method (50 ± 7.4 nm)(SD) was close to the distribution given by the manufacturer and was in the range of what has been described for NTA or DLS (Dynamic light scattering) [24] . Further determination of the particle size relies mostly on the evaluation of the scattered signal values: indeed, being mostly in the Rayleigh scattering regime the scattered *amplitude* for quasi-spherical particles is proportional to the third power of the diameter [16].

We then counted the mean number of particles per frame of observed signals for each dilution of 50 nm particles in $10^{-8}$ mL (the standard volume of measurement). We observed a satisfactory correlation between the number of detected signals and particle dilutions **(Fig. S2C)**.. Finally, by comparing the average number of particles per frame from independent videos of the same sample we estimated the variability between movies to be around 10%.

These initial results demonstrated the validity of our method for detecting and quantifying nanometer particles and encouraged us to analyze suspensions of known viruses.

## 3. Results using homogeneous viral suspensions of different sizes, shapes and genetic composition

To test the usefulness of our method to detect viruses, we first chose representatives of virus families present in aquatic environments such as T4 phage (for Myoviruses) λ phage (for Siphoviruses), and T7 phage (for Podoviruses). These viruses belong to the order Caudovirales [30]. Myoviruses and podoviruses possess, in addition, fibers at the tip of the tail. We used also poliovirus as a representative of the Picornaviruses (RNA virus) as well as a filamentous ssDNA virus M13. Our aim with this collection of viruses was to evaluate a variety of shapes and genetic materials. The results presented in **Table 1** show that we can detect viruses according to the size of their capsid (or head) and possibly their shape and independently of their genetic material, be it dsDNA, RNA or ssDNA.

Some viruses such as M13 phage or plant viruses possess elongated flexible or rigid helical capsids. To determine the scattering signal values of viruses with this type of morphology, we analyzed a lysate of M13 phages. We observed maximum signal values of 12.6± 3 a.u (SD), which is close to what we observed with poliovirus particles. This maximum signal corresponds to the case of a cylinder oriented perpendicular to the optical axis; because the cylinder length exceeds the diffraction spot diameter we can compute its scattering cross-section as an infinite cylinder limited to the diffraction spot diameter [16]. We found that the cross-section was close to that of a 30 nm spherical particle diameter [16]. Interestingly the diffraction spot was not strictly circular suggesting a way to identify helical capsids or other helical structures such as bacterial flagella.

Table 1: Morphological properties of known viruses and sizes determination by interferometry

|  | Head diameter Tail length (nm) | Scattering signal (a.u) | Diameter$_{scat}$ ±SD | Diameter$_{BM}$ ±SD |
|---|---|---|---|---|
| T4 phage | 111x78 113x16 | 151.2±45.7 | 85.5±9.5 | 86.2±6.7 |
| Lambda phage | 60 150x8 | 53.9±20.6 | 63.2±11.2 | 55±12 |
| T7 phage | 60 17x8 | 46.1±13.1 | 56.6±8.2 | 52±14.5 |
| 50nm beads | 50 | 30±10.5 | 50±7.4 | 54.1±17 |
| Polio virus | 35 | 10.5±2.9 | 35±3.2 | 37±15.9 |
| M13 phage | 800x7.5 | 12.6±3 | Not determined | Not determined |

## 4. Brownian motion analysis: diameter measurements and evidence for anisotropic trajectories

Viruses have no means of locomotion; they rely entirely on mass diffusion and their mobility depends on parameters such as size and shape. Our method tracks particles trajectories thus we can also determine the virus diameters by measuring the average jumps between two successive frames for each particle. This step deals with reconstruction of 2D trajectories from a scrambled list of localized particle coordinates determined at discrete times (in consecutive video frames)[31].

We plotted the diameter of the 50 nm beads by the two measurements, scattering signal (diameter$_{scat}$ proportional to the refractive index difference Δn with water refractive index Δn =1.5-1.33 for viruses) and average displacement (diameter$_{BM}$) (**Table1, Fig. S2D**). However, we noticed a larger dispersion when diameters were computed with Brownian displacement (38-68 nm) than with the scattering signals (about 60% of particles in the 45-55 nm range). Indeed, the particle jump between successive images is proportional to the inverse of the square root of the diameter whereas the scattering signal varies with the third power of the diameter. Despite of this dispersion it was of interest to compare the diameter distributions obtained through the scattering level and through the Brownian motion. In particular we reasoned that it could reveal the presence of nanoparticles of different refractive indexes because the slope of the curve diameter$_{sca}$ vs diameter$_{BM}$ is proportional to $\Delta n$ (see below).

To improve discrimination between viruses of different families we also analyzed their Brownian motion trajectories. The trajectories of Lambda phage particles showed the typical Brownian trajectories of spherical particles). We also found anomalous Brownian motion trajectories for the T4 phage which exhibited a marked tendency to preferentially move in a forward direction, in the sense that they made more anisotropic jumps than the other viruses such as Lambda phage that exhibited rather symmetric random walks. To characterize the type of trajectories (as commonly used for photon scattering; http://omlc.org/classroom/ece532/class3/gdefinition.html) we measured the average cosine of the angle between two successive jumps of the same particle and determined a coefficient of anisotropy for the trajectories. More precisely if we call $\theta_1, \theta_2, \theta_3$......the successive angles that the particle trajectory experiences after each step, for an isotropic diffusion the anisotropic factor $g=<\cos(\theta)>=0$ and g will get a value $0<g<1$ for an anisotropic diffusion. In order to establish after how many steps the anisotropic particles trajectory behaves as an isotropic trajectory one has to compute $<\cos(\theta_1+\theta_2 + \theta_3......\theta_n)>$ and consider when this mean value reaches our experimental errors range. Due to the randomness of independent events one can write:

$<\cos(\theta_1+\theta_2)> = <\cos(\theta_1)><\cos(\theta_2)> - <\sin(\theta_1)><\sin(\theta_2)> =<\cos(\theta)>^2 - <\sin(\theta)>^2 = g^2$ because $<\sin(\theta)>= 0$ for symmetry reasons. One can then shows that $<\cos(\theta_1+\theta_2 + \theta_3......\theta_n)>=g^n$

One can thus compute the number n of steps to get $g^n$ smaller than our experimental errors. On the T4 trajectory, the error is 0.028, so seven steps are necessary to reach a non measurable anisotropy. Seven steps correspond to an observation time of 50ms. Lambda phage particles show an anisotropic coefficient with low values that correspond to classical Brownian trajectories unlike T4 phage, which exhibited a high value of anisotropic coefficient in agreement with the described anisotropic trajectories (**Table 2** ). No evidence of anisotropic behavior was found for the other analyzed viruses.

Table 2. Coefficient of anisotropy for Brownian trajectories (computed with trajectories of at least 10 jumps)

|  | Number of analyzed trajectories | Coefficient of anisotropy |
|---|---|---|
| T4 phage | 453 | 0.61±0.03 |
| Lambda phage | 661 | 0.063±0.002 |

Few studies have described the Brownian motion of anisotropic macromolecules, but of particular interest is the study of Han et al (2006) [32] on ellipsoidal particles. But here we have to face more complex geometrical shapes. The T4 bacteriophage like all myoviruses possesses a capsid and a tail prolonged with long tail fibers. It has been suggested that instead of random displacement of the phage, the fibers guide the phage to the bacteria suggesting oriented trajectories in agreement with what we observed [33].

Our results demonstrate that our interferometric detection method coupled with Brownian motion analysis can resolve viruses of various sizes and identify a specific signature for myoviruses.

## 5. Analysis of vesicles secreted by the diatom *Phaedactylum tricornutum*

All eukaryotes produce vesicles we thus analyzed membrane vesicles produced by the diatom *P. tricornutum*. Culture medium was filtered on 0.22μm and concentrated on Amicon membrane (Millipore, cut off 30kDa). We first plotted the diameter of each particle computed with its scattering signal as a function of the diameter computed with its average displacement (**Fig. 3A**). This type of analysis is unbiased, as it has no prejudice towards the type of nanoparticles present in the sample. Interestingly, we observed that the particles were not distributed on the bisector (n=1.5), as expected for viruses, but on lines of lower refractive indexes, suggesting a lower level of dry matter in these particles than in viruses. The clustering analysis of diameters determined by Brownian motion showed several populations from 40 nm up to 350 nm (analysis performed with Mclust package of the R software) [34]. From the measurements of the scattered signals we can conclude that the dry matter of the vesicles that determine the refractive index (soluble proteins as well as membrane proteins) is mainly the same whatever their size: indeed for all the recorded diameters the scattered signal (that reflects the dry matter level of each vesicle or the difference of its refractive index with water) tends to retain the same values.

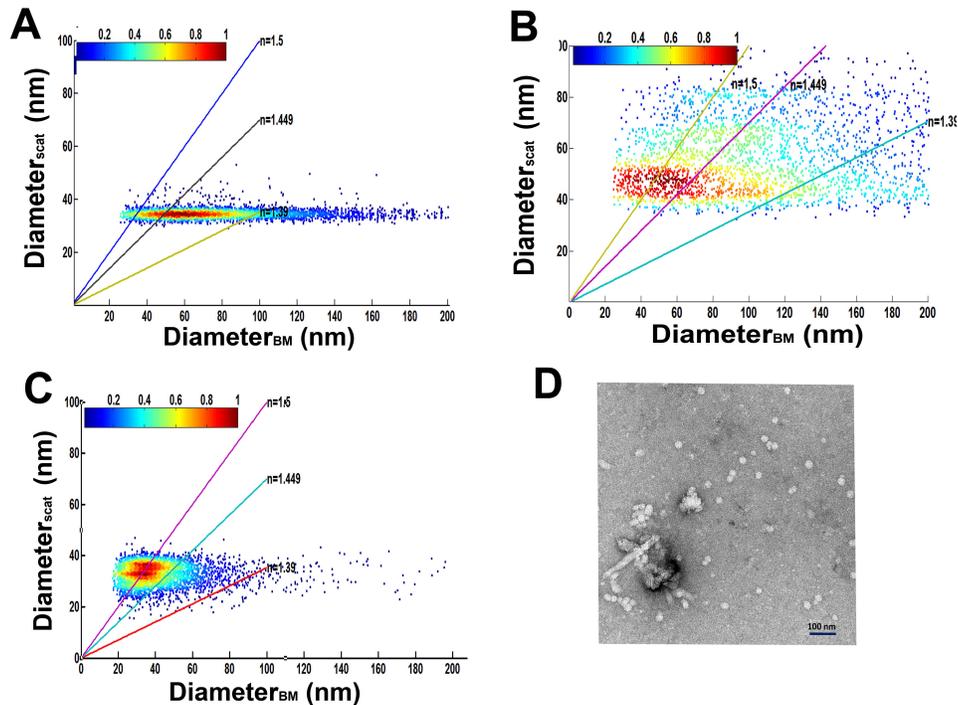

Fig. 3. Analysis of membrane vesicles and *Tara*-oceans samples. A) Plot of diameters (N= 4,788) computed with scattered signals function of diameters computed by Brownian motion of membrane vesicles secreted by *P. tricornutum*. The line of slope 1 corresponds to a refractive index of 1.5. The colored bar corresponds to the number of particles per unit area. B) Analysis of *Tara* Oceans samples from the Dalmatian coast (Station 23) Plot of the particle diameters (N=4,247) computed with their scattered signals and Brownian motion. Lines indicate the refractive indexes of the different particles (viruses and vesicles). The colored bar corresponds to the number of particles per unit area (for A and B). We intentionally threshold the Diameter$_{BM}$ to 20nm, corresponding to our detection noise. C) Analysis of *Tara* Oceans sample from an oligotrophic open ocean region (Station 98). Plot of the particle diameters (N=8,136) computed with their scattered signals and Brownian motion. Lines indicate the refractive indexes of the two types of particles, small viruses and vesicles. D) : Electron micrograph of viral fraction from *Tara* Oceans station 98 (concentrated 200 times) showing abundance of small particles. Scale bar: 100 nm.

We thus conclude that we can detect secreted vesicles, moreover, our analysis suggests that we could differentiate between non-enveloped viruses and vesicles by combining diameters computed with scattering signals and average displacements because viruses exhibit diameters aligned with a refractive index n=1.5 in our representation, while vesicles are aligned with lines corresponding to lower refractive indexes. We therefore went on to use our method to detect biotic nanoparticles abundant in marine environments.

## 6. Contrasting compositions of nanoparticles between coastal and oligotrophic marine environments

We examined two samples from the *Tara* Oceans circumnavigation [25,26], a coastal sample (from Dalmatian coast, Croatia in the Mediterranean Sea; Station 23) and a sample from an oligotrophic region of the Pacific Ocean (near Easter Island; Station 98). Samples were filtered on 0.22μm, concentrated or not and analyzed.

Our analysis of the sample from the Dalmatian coast (station 23, $10^7$ particles/mL)(**Fig. 3B**) predicted 17% of virus particles suggesting that a large fraction of this sample was composed of vesicles. The statistical analysis of $diameter_{scat}$ of this fraction shows 2 classes of viruses with averages of 46 and 67 nm in diameter that were confirmed by the analysis of $diameter_{BM}$ (analysis performed with Mclust package of the R software) [34]. Further more the distribution of sizes was in the same range as the distribution of diameters observed by qTEM with different fractions of the same station [35]. To sort the types of viruses in this sample, we analyzed the trajectories of particles that experienced at least 10 jumps (787 particles). We considered two groups, smaller and larger than 60 nm in diameter and calculated their coefficients of anisotropy. We observed that 60% of particles (256 particles) in the group with diameters larger than 60 nm (17% in the group of diameter less than 60 nm) exhibited a coefficient of anisotropy larger than 0.50 (0.54±0.2 (SD)) suggesting that this group is enriched in myoviruses. Interestingly, we estimated that about 20% of all analyzed trajectories corresponded to myovirus trajectories, which is in agreement with the results observed by qTEM [35]. The rest of the particles for which no correlation was observed between the two measurements of diameter corresponds to particles with lower refractive indexes and are likely vesicles from 80 nm to 200 nm in diameter and over.

The analysis of a sample from Station 98 ($0.9 \ 10^8$ particles/mL), which is from an oligotrophic environment showed that more than 50% of the particles were predicted to be viruses (**Fig. 3C**). Analysis of viral fraction predicted a majority of viruses centered on 40nm in diameter (about 30%) (analysis performed with Mclust package of the R software) [34]. TEM of the same sample confirmed the abundance of small particles (**Fig. 3D**; mean= 35.3±6.8 nm (SD)). Finally, less than 10% of predicted viral fraction corresponded to viruses larger than 60nm. These results are in agreement with the low percentage of "classical" phages described by qTEM in this station [35]. The remaining nano-particles identified in Station 98 could correspond to vesicles about 60 nm in diameter or larger.

In summary, from these two environmental samples we were able to show that in the coastal sample we observed an abundance of vesicles while an oligotrophic environment contained fewer vesicles but exhibited an abundance of small viruses.

## 7. Discussion

We show here a new sensitive and stable common path full-field interferometric method to detect and count viruses and vesicles that is fast and easy to implement. Our set up based on LEDs avoids using lasers and the associated speckle induced by unwanted stray light. In addition, shot noise limited signals can be detected using inexpensive megapixel fast cameras. With this approach it is possible to analyze convection-free very small sample volumes (we successfully tested 0.5 μL) and we have been able to enumerate viruses in the 30-100 nm diameter ranges through their amount of scattered light coupled with analysis of their Brownian motion trajectories. Furthermore we revealed for the first time the anisotropic Brownian motions of T4 virus particles and introduced a parameter to quantify the anisotropy (**Table 2**). Anisotropy measurement was made possible though the use of a 150 fps camera that is unlikely to be possible with the NTA 30 fps camera. To the best of our knowledge, these Brownian motions have not been detected prior to this study. We forecast therefore that our results will stimulate numerical models of the hydrodynamic properties of complex structures like the different viruses studied here. Interestingly we show that myoviruses of different sizes from environmental samples exhibit the same type of complex Brownian motion as purified T4 viruses. Furthermore, we show that we can detect membrane vesicles with our method and differentiate them from viruses (**Fig.3).**

The method described here can also be applied to samples from a variety of environments such as soil or gut. Finally, it will be highly interesting to improve the sensitivity and the speed of detection because it will open the way to the analysis of even smaller particles such as RNA viruses that have been proposed to play major roles in some oceanic environments [13]. In this direction we expect the technique to benefit from higher power LEDs and new cameras with higher full-well capacities and frame rates. In addition, with such cameras we should be able to increase the specificity of our detection through the analysis of rotational Brownian motions, taking advantage of the polarization properties of the scattered light. We expect thus to reveal the anisotropy of the scattered light for virus structures that we missed due to camera speed and our LED power limitation. Finally, in this study, the small sample volume was filled manually with a micropipette. For the continuous monitoring of a large variety of samples, a microfluidic channel with an oil separation between two successive micro volumes would be well adapted for an automatic analysis [36] . Such a device could also be used for sorting viruses and vesicles.

**Aknowlegements:** We would like to thank all the persons who helped us to establish the method: Benoit Quenet (ESPCI-ParisTech trainee), Charles Brossolet (LLTech), Dr Van Etten and J. Gurnon (Nebraska Wesleyan University), Pr G. Sesonoff (UPMC, Paris), Dr H. Lecoq (INRA, Avignon), Dr M. Dreyfus, Dr E. Hajnsdorf and Dr J. Plumbridge (IBPC Paris). We deeply acknowledge Paris Science et Lettres for initial financial support and the Fondation Pierre-Gilles de Gennes for strong financial support. We thank Dr J-M Guigner for help with electron microscopy. Thanks also to Dr C. Thery and J Kowal (Institut Curie) for the nanosight test. Thanks to S. Brouillet for her help with R software and to E. Cailleux for his help with the figures. The samples from *Tara* Oceans Station 23 and Station 98 were provided by Dr Colomban de Vargas, and we are proud to contribute to this great human and scientific adventure. This article is contribution number XXXX of the *Tara* Oceans Expedition 2009–2012.

**SUPPLEMENTARY FIGURES AND LEGENDS**

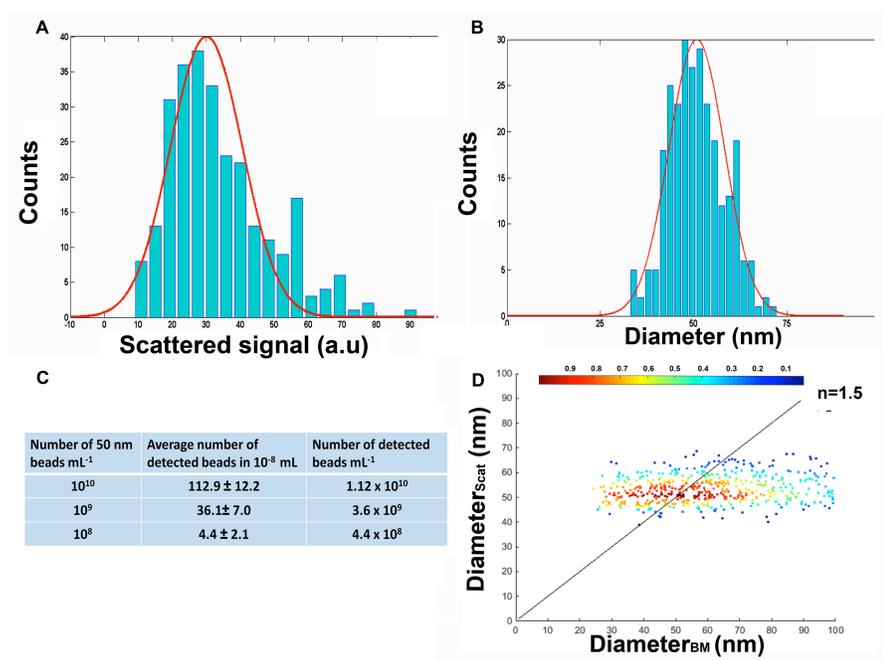

Fig. S1. Calibration with 50 nm silica beads. **A)** Histogram of scattered signals observed with beads (concentration: $10^{10}$ particles/mL). **B)** Histogram of particle sizes. Note that the uncertainly corresponds to 7%, while it is about 30% in A). **C)** Correlation between the number of particles measured and dilutions. **D)** Beads diameters measured through the scattering level as function of their diameters measured through the Brownian motion (N=606). As expected (see text) the dispersion of the measurements is much larger in this later case. We considered particles that experienced at least 10 measured jumps. The colored bar corresponds to the number of particles per unit area.